\documentclass[aps,prl,twocolumn,showpacs,superscriptaddress]{revtex4-1}
\usepackage[latin1]{inputenc}
\usepackage{graphicx}
\usepackage{epstopdf}
\usepackage{epsfig}
\usepackage{xfrac}
\usepackage{indentfirst}
\usepackage{amsmath, amsthm, amssymb} 
\usepackage{bm}
\input epsf
\usepackage{float}
\usepackage[colorlinks=true, a4paper=true, pdfstartview=FitV, linkcolor=blue, citecolor=blue, urlcolor=blue]{hyperref}
\usepackage{braket}
\begin{document}

\title{Equation of state of a laser cooled gas}

\author{J. D. Rodrigues}
\affiliation{Instituto de Plasmas e Fus\~{a}o Nuclear, Instituto Superior T\'{e}cnico, Universidade de Lisboa, 1049-001 Lisbon, Portugal}
\author{J. A. Rodrigues}
\affiliation{Departamento de F\'isica, Faculdade de Ci\^{e}ncias e Tecnologia, Universidade do Algarve, Faro, Portugal}
\affiliation{Instituto de Plasmas e Fus\~{a}o Nuclear, Instituto Superior T\'{e}cnico, Universidade de Lisboa, 1049-001 Lisbon, Portugal}
\author{O. L. Moreira}
\affiliation{Instituto de Plasmas e Fus\~{a}o Nuclear, Instituto Superior T\'{e}cnico, Universidade de Lisboa, 1049-001 Lisbon, Portugal}
\author{H. Ter\c cas}
\affiliation{Physics of Information Group, Instituto de Telecomunica\c c\~{o}es, Lisbon, Portugal}
\author{J. T. Mendon\c ca}
\affiliation{Instituto de Plasmas e Fus\~{a}o Nuclear, Instituto Superior T\'{e}cnico, Universidade de Lisboa, 1049-001 Lisbon, Portugal}


\begin{abstract}
\begin{footnotesize}

We experimentally determine the equation of state of a laser cooled gas. By employing the Lane-Emden formalism, widely used in astrophysics, we derive the equilibrium atomic profiles in large magneto optical traps where the thermodynamic effects are cast in a polytropic equation of state. The effects of multiple scattering of light are included, which results in a generalized Lane-Emden equation for the atomic profiles. A detailed experimental investigation reveals an excellent agreement with the model, with a two-fold significance. In one hand, we can infer on the details of the equation of state of the system, from an ideal gas to a correlated phase due to an effective electrical charge for the atoms, which is accurately described by a microscopical description of the effective electrostatic interaction. On the other hand, we are able map the effects of multiple scattering onto directly controllable experimental variables, which paves the way to subsequent experimental investigations of this collective interaction.

\end{footnotesize}
\end{abstract}

\maketitle

\par
The concept of equation of state, the relationship between thermodynamic state variables such as pressure, $P$, temperature, $T$, and volume, $V$, has evolved beyond the original formulation of Clapeyron for an ideal gas, $PV = N K_B T$. In particular, it has been argued that there is a universal form for the equation of state for solids \cite{eq_state_1}. Matter at nuclear density may also allow for a description in terms of an equation of state, with implications on astrophysical observations of neutron stars \cite{eq_state_2, eq_state_3, eq_state_4}. In cosmology, an equation of state is expressed in terms of the ratio of pressure $P$ to energy density $\rho$ \cite{eq_state_5}, and can be measured from observations on supernova distances \cite{eq_state_6}. Even Einstein's equation of general relativity can be derived from the Gibb's relation $\delta Q = T dS$, with $\delta Q$ being the energy flux and $T$ the Unruh temperature, allowing for the interpretation of Einstein's equation as an equation of state \cite{eq_state_7}. More recently, with the advent of laser cooling, attention has been given to the thermodynamics of degenerate Bose \cite{eos1, eos2} and Fermi gases \cite{eos3}. 

\par
In astrophysics, the equation-of-state plays a central role in the study of stellar structure, where hydrostatic equilibrium condition of a polytropic gas under the gravitational field and thermodynamic pressure leads to the celebrated Lane-Emden equation \cite{Lane1, Lane2}. Based on this approach, Chandrasekhar derived the mass limit for a white dwarf \cite{Lane3}. Moreover, the Lane-Emden formalism has also been used to test alternative theories of gravity in stars beyond the standard models of stellar structure \cite{Lane5}. 

\par
In this Letter, we extend the Lane-Emden formalism to experimentally determine the equation of state of a laser cooled gas, by directly measuring the atomic density profiles of large magneto-optical traps. In our experiment, the hydrostatic equilibrium condition is provided (in first order) by the balance between the harmonic confinement - the analog of the gravitational force - and the thermodynamic pressure, cast in the form of a polytropic equation of state. However, when the cooling lasers tuned close to the atomic resonance, multiple scattering of light occurs and an additional collective interaction appears, due to the exchange of scattered photons with nearby atoms \cite{sesko_2, dalibard}. In this regime, the atoms experience a Coulomb-like long-range interaction \cite{Q}, therefore allowing to regard the system as an effective one-component trapped plasma \cite{livro}. Thus, the condition of hydrostatic equilibrium results in a generalized Lane-Emden equation, which encompasses the joint effects of harmonic confinement, thermodynamics and radiation pressure due to multiple scattering of light. Here, we provide the experimental evidence of the multiple scattering of light in the equation of state of a laser cooled gas and extracting the polytropic exponent by fitting the density profiles to our theory. A microscopical description of the interaction induced by multiple scattering is introduced to explain the experimental deviations from the ideal gas.

\par
A fluid description of a laser-cooled gas confined in a magneto-optical trap (MOT) may be introduced with the usual continuity and Navier-Stokes equations \cite{tito_2008, livro, hugo}
\begin{equation} \label{eq:continuity}
\frac{\partial n}{\partial t} + \boldsymbol{\nabla} \cdot (n \boldsymbol{v} ) = 0, 
\end{equation}
\begin{equation} \label{eq:Nav-Stokes}
\frac{\partial \boldsymbol{v}}{\partial t} + \boldsymbol{v} \cdot \boldsymbol{\nabla} \boldsymbol{v} = -\frac{ \boldsymbol{\nabla} P}{m n} + \frac{\boldsymbol{F}_{t}}{m} + \frac{\boldsymbol{F}_{c}}{m},
\end{equation}
where $n$ and $\boldsymbol{v}$ represent the gas density and velocity field, respectively, and $m$ is the atomic mass. The collective interaction due to multiple scattering of light is determined by a Poisson-like equation

\begin{equation} \label{eq:lap1}
\boldsymbol{\nabla} \cdot \boldsymbol{F}_{c} = Qn, 
\end{equation}
where $Q = (\sigma_R - \sigma_L)\sigma_L I_0 / c$ represents the square of the effective charge of the atoms \cite{Q}, $I_0$ is the total intensity of the beams and $c$ is the speed of light. Here, $\sigma_R$ and $\sigma_L$ represent the emission and absorption cross sections, respectively \cite{walker_1990}. The term $\boldsymbol{F}_{t}$ encompasses the usual MOT cooling and restoring force as determined by $\boldsymbol{F}_{t} = - \alpha \boldsymbol{v} - \kappa \boldsymbol{r}$, corresponding to a damped-harmonic oscillator. The trapping potential is then assumed to be harmonic, which is a reasonable assumption for the anti-Helmholtz coils configuration. The spring constant is approximately given by $\kappa = \alpha \mu_{B} \nabla B / \hbar k = \kappa(\delta, I_0, I_s)$ where $\mu_B$ represents the Bohr magneton, $\alpha = \alpha (\delta, I_0, I_s)$ the friction coefficient from Doppler cooling, $\delta$ is the laser detuning and $I_s$ is the atomic saturation intensity. For magnetic fields gradients of the order $\sim 10$ G/cm, deviations from the harmonic case are obtained for very large traps, namely $R \sim 1$ cm. In our experiments, we use $R\sim 2$ mm, so we may neglect anharmonic effects in the confinement. The thermodynamic effects are cast in the form of a polytropic equation of state
\begin{equation}
P(r) = C_{\gamma} n(r)^{\gamma},
\end{equation}
where $\gamma$ is the polytropic exponent, $C_{\gamma} = P(0) / n(0)^{\gamma}$ is a constant determined by the thermodynamic properties of the system, and $P(0)$ and $n(0)$ the pressure and density at the center of the cloud. Introducing the condition of hydrostatic equilibrium in Eqs. (\ref{eq:continuity}), (\ref{eq:Nav-Stokes}) and (\ref{eq:lap1}) yields the generalized Lane-Emden equation \cite{hugo}
\begin{equation}\label{eq:Lane_eq}
\gamma \frac{1}{\zeta^2} \frac{d}{d \zeta} \left( \zeta^2 \theta^{\gamma -2} \frac{d \theta}{d \zeta} \right) - \Omega \theta  +1 = 0,
\end{equation}
where $n(r) = n(0)\theta(r)$ and $r = a_{\gamma} \zeta$, with $a_{\gamma} = \sqrt{\frac{C_{\gamma}}{3 m \omega_0^2}} n(0)^{(\gamma-1)/2}$ a typical scale in the system and $\omega_0^2 = \kappa / m$ the trap frequency. The dimensionless parameter $\Omega = Qn(0)/3m \omega_0^2 = \omega_p^2 / 3 \omega_0^2$, with $\omega_p = \sqrt{Q n(0) / m}$ the equivalent plasma frequency \cite{tito_2008}, is the of ratio multiple scattering (plasma) to the trapping forces. The stability of the solutions can therefore be directly related to $\Omega$, with stable solutions existing for $0 \leq \Omega < 1$, as confirmed both numerically as by linear stability analysis \cite{hugo}. The two limits correspond to distinct physical relevant scenarios. In one hand, for smaller traps with $10^7 \sim 10^8$ atoms and large laser detuning, the dynamics is determined by the thermal effects and multiple scattering is negligible, corresponding to the limit when the atoms have no effective charge, $Q\rightarrow0$, or equivalently, $\Omega \rightarrow 0$. In this limit, the equilibrium density profiles, for a spherically symmetric cloud, are given by 
\begin{equation}
n(r) = n(0) \left(1-\frac{\gamma-1}{6\gamma}\frac{r^2}{a_\gamma^2} \right)^{1/(\gamma-1)}.
\end{equation}
The case $\gamma = 1$ and $C_1=k_B T$ (isothermal gas) simply corresponds to the Maxwell-Boltzmann equilibrium
\begin{equation}  \label{eq:exp_law}
n(r) = n(0)e^{-U(r)/k_B T} = n(0)e^{-r^2/R^2},
\end{equation}
with $R=\sqrt{2 k_B T / m \omega_0^2}$ the $1/e$ radius of the cloud. Later we shall see that the isothermal case is the most relevant solution in this limit. On the other hand, for very large traps with $N \gtrsim 10^8$ atoms and small detuning, $\vert \delta \vert \lesssim \Gamma$, with $\Gamma$ being the linewidth of the transition, the process of multiple scattering dominates, $\Omega \rightarrow 1$, and thermal effects can be ignored. In this case, a qualitative analytical solution can be found setting $\gamma \rightarrow 0$, yielding a step-like profile
\begin{equation}
n(r) = n(0) \Theta(r-R),
\end{equation}
with $n(0) = 3m\omega_0^2/Q$, $R=\left(\frac{3N}{4\pi n_0}\right)^{1/3}$ the radius of the cloud and $\Theta(r-R)$ the Heaviside function. These two limiting cases - the temperature-limited and multiple-scattering regime, respectively - are well known and have been reported by the early experiments \cite{sesko_2, townsend}, although no relation with an equation of state has been established so far. Our experiments provide a quantitative measurement of the intermediate regimes, both theoretically and experimentally, on the equation of state of the gas, and its dependence on the effective charge  $Q$. We shall also mention that a third regime may be possible, namely the two-component regime \cite{townsend}. In this case, there is a strong confinement near the center of the cloud, due to the influence of the magnetic field on the optical pumping between the Zeeman sublevels of the ground state, and a weaker confinement in the outer region due to the Zeeman shift of the various excited sublevels. However, in large traps, almost all the atoms occupy the weak confinement volume, the trap dynamics is essentially the one presented here and the presence of polarization gradients in the laser fields no longer influences the behaviour the of system. The influence of this regime is thus safely excluded both from our theoretical model and experimental analysis. 

\par
We now perform a detailed investigation of the equilibrium atomic profiles in a cold trap. Our experimental apparatus consists of a MOT \cite{MOT}, where $^{85}$Rb atoms are collected from a dilute vapour in a background pressure of $\sim 10^{-8}$ Torr. Six independent trapping (and cooling) laser beams cross the center of the trap with beam waist of $w\sim4$ cm, power per beam $P \sim 40$ mW and wavelength $\lambda\sim 780$ nm. The beams are not retro-reflected, thus avoiding feedback instability mechanisms \cite{MOT_instability}. The trapping laser operates on the D2 line of $^{85}$Rb ($F=3 \rightarrow F'=4$), and is red-detuned by $\delta$, which can be precisely controlled by a double passage through an acousto-optic modulator (AOM). The transition linewidth is approximately $\Gamma/2\pi = 6$ $MHz$. A magnetic field gradient ($\nabla B$) created with a pair of water-cooled coils in an anti-Helmholtz configuration (zero field in the center of the trap) generates a spatially dependent Zeeman split of the energy levels, yielding the restoring force of the trap. An additional repump beam, operating on the hyperfine levels $F=2 \rightarrow F'=3$ of the $D2$ line repopulates the trapping transition. The repump detuning is set by searching for the maximum fluorescence signal, corresponding the larger number of atoms in the trap. We thus obtain a cold cloud with $T\sim 100$ $\mu K$ and $N \sim 10^7$ to $N \sim 10^{10}$ atoms, depending on the laser detuning $\delta$. A CCD camera collects the fluorescence signal, illuminating the cloud with far from resonance light ($\delta = - 4 \Gamma$), to avoid multiple scattering during the imaging process. In this way we measure the atomic distribution of the trap, integrated along the line-of-sight of the camera. Two additional CCD cameras, positioned in orthogonal directions, allow us to monitor the shape of the cloud. By using half-wave plates we independently control the intensity of the six trapping beams to achieve a spherically symmetric atomic ensemble.

\par
\begin{figure}[htp]
\centering
\begin{tabular}{c}
\includegraphics[width=86mm]{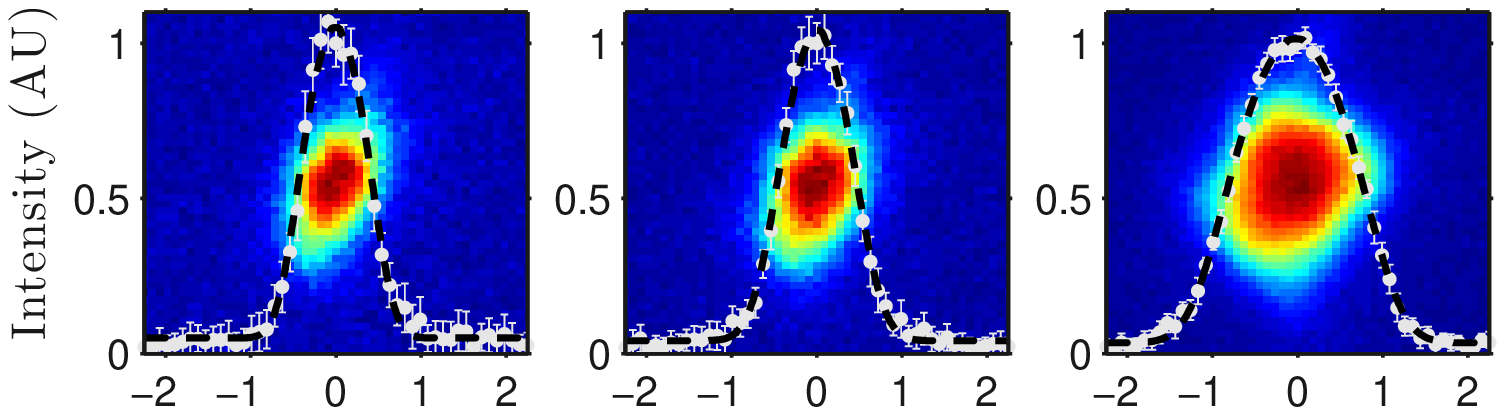}\\
\includegraphics[width=86mm]{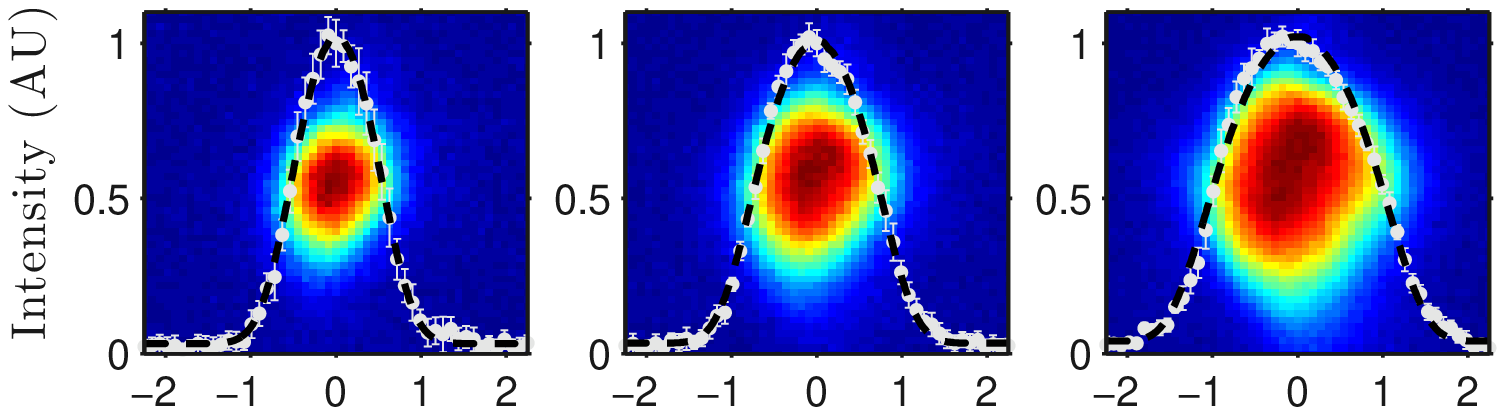}\\
\includegraphics[width=86mm]{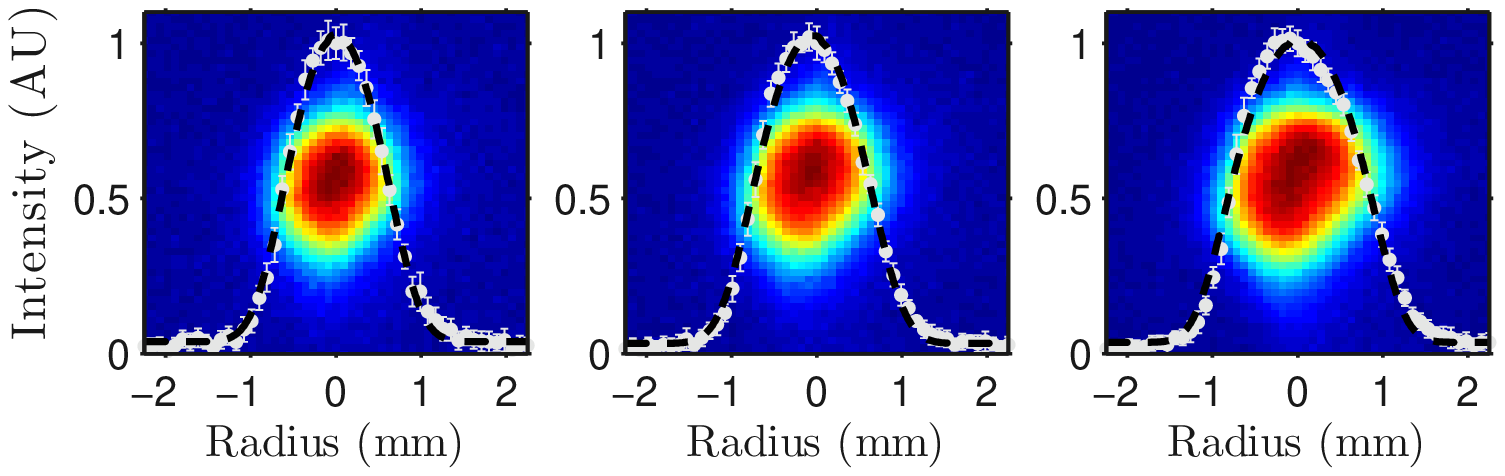}
\end{tabular}
\caption{(color online) Integrated atomic density profiles (in false color code), and a one-dimensional horizontal cut passing through the center of the cloud (white points). The corresponding fitting curves to numerically generated solutions of the Lane-Emden equation in (\ref{eq:Lane_eq}) are also displayed (black dashed line). The magnetic field gradient varies from top to bottom as $\nabla B=7.5$, $10$, and $12.5$ G/cm. From left to right, we vary the detuning as $\delta = -3.2$, $-2.8$ and $-2.4$ $\Gamma$. The error bars here correspond to standard deviations of the CCD pixel values, taken over 30 realizations.}
\label{fig_density}
\end{figure}

For each experimental condition, determined by the laser detuning $\delta$ and magnetic field gradient $\nabla B$, we average the experimental CCD profiles over 30 realizations. One-dimensional profiles are obtained by cutting through a direction crossing the center of the cloud, whose coordinates are defined as the ``center-of-mass'' of the two-dimensional image. Each experimental profile is fitted with the general solution of Eq. (\ref{eq:Lane_eq}), numerically computed with a fourth order explicit Runge-Kutta method, and integrated along one arbitrary direction (our system is spherically symmetric). The agreement between the experimental data and the theoretical model is excellent, for the whole range of experimental parameters investigated here - see Fig. (\ref{fig_density}). By deacreasing the laser detuning, i.e. approaching the resonance, we clearly observe a transition from a Gaussian to a flattened density profile, also known as {\it water-bag} profile (which corresponds to a paraboloidal curve, when integrated along the line-of-sight). It corresponds to a crossover from the temperature-limited regime, with $\gamma \sim 1$ and $\Omega \sim 0$, to the multiple scattering regime, with $\Omega \rightarrow1$.
\par
\begin{figure}
\centering
\includegraphics[scale=0.68]{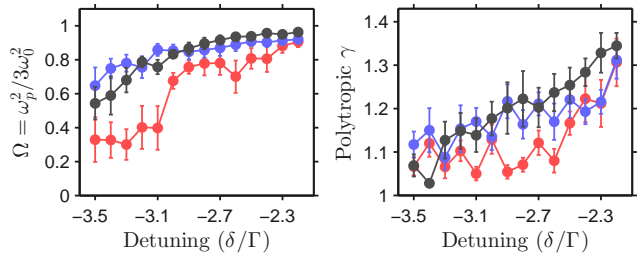}
\caption{(color online) In the left panel, we plot the normalized plasma frequency $\Omega = \omega_p^2 / 3 \omega_0^2 $ as a function of detuning $\delta$. The right panel depicts the polytropic exponent $\gamma$, also as a function of detuning $\delta$. The different curves correspond to $\nabla B = 7.5$ G/cm (red line), $\nabla B = 10$ G/cm (black line) and $\nabla B = 12.5$ G/cm (blue line).}
\label{fig_Omega}
\end{figure}

In what follows, we determine how the physical quantities of the model, the dimensionless plasma frequency $\Omega$ and the polytropic exponent $\gamma$, scale explicitly as a function of the experimental parameters along the crossover. For that purpose, we take several averaged realizations of the 2D profile from which one-dimensional profiles are extracted. By numerically fitting the latter to the general solution of Eq. (\ref{eq:Lane_eq}), we plot $\Omega$ and $\gamma$ against $\delta$ and $\nabla B$. The error bars thus correspond to the statistical standard deviations of a large number of fitted profiles, as depicted in Fig. (\ref{fig_Omega}). Note that we are not attributing any error to the detuning $\delta$, which is stable in our set-up. On the other hand, the laser power is not constant, with a $5\sim 10$ $\%$ drift that we can not overcome. This may indeed induce some fluctuations in the fitted parameters, although its significance should be mitigated by using large statistics. The size of the statistical error bars do, in fact, reflect the experimental ``jitter" associated with not only the laser power fluctuations, but other parameters drifts.
\par
As expected, when working closer to resonance, the effect of multiple scattering becomes more important, not only because of the higher value of the cross sections $\sigma_R$ and $\sigma_L$, but also because the number of atoms in the trap also grows, increasing the probability of a photon being reabsorbed before leaving the system. At the same time, larger deviations from the ideal gas ($\gamma = 1$) are also observed, as the gas starts to behave like a (weakly) coupled one-component plasma. Very close to resonances, $\vert \delta \vert \simeq 2 \Gamma$, we observe the onset of some mechanical instabilities in the trap, characterised by an oscillatory behaviour of the fluorescence signal. This effect has been observed by other authors and reported in the literature as a self-sustained instability \cite{Robin}, which are related with a competition between the confining force of the trap and the increasing repulsive interaction associated with multiple scattering. As our model relies on dynamically stable regimes, we excluded data taken for $\vert \delta\vert  <2 \Gamma$.
\par
\begin{figure}
\centering
\includegraphics[scale=0.68]{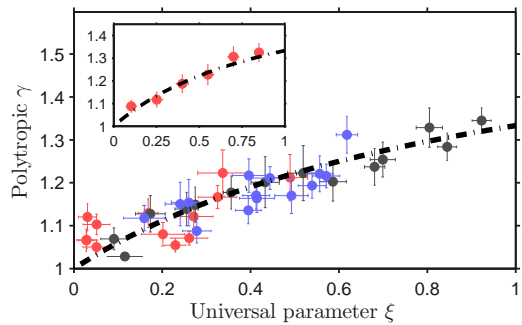}
\caption{(color online) Polytropic exponent, $\gamma$, as a function of the universal parameter $\xi$, for $\nabla B = 7.5$ $G/cm$ (red points), $\nabla B = 10$ G/cm (black points) and $\nabla B = 12.5$ G/cm (blue points). All points corresponding to different experimental conditions fall on the universal curve defined by Eq. (\ref{eq:correction}). The inset corresponds to the same experimental points but binned in the $\xi$ parameter and equally spaced.}
\label{fig_correction}
\end{figure}

At this point, it is pertinent to query about the dependence of the equation of state, as determined by the the polytropic exponent $\gamma$, on the effective charge of the atoms. To unravel this dependence, we develop a simple microscopical theory for the interactions in the system. We start by explicitly computing the total energy, $U = U_T + U_C$, with $U_T = \tfrac{3}{2}NK_BT$ the usual thermal energy and $U_C$ the energy associated with the Coulomb interactions and responsible for the deviations from the ideal gas. To compute $U_C$ we start with Poisson equation, $\nabla^2 \phi (r) = -\tfrac{q}{\epsilon_0} n(r)$, with $q = \sqrt{\epsilon_0 Q}$ the effective charge of the atoms and $n(r) = n(0)$ assumed to be constant throughout the cloud - remember the water-bag solution - which allows us to explicitly derive an analytical solution. After determining the electrostatic energy of the system, we can derive the corresponding pressure resulting from the exchange of scattered photons as $P_C = \tfrac{Qn(0)^2}{6}\left( \tfrac{3V}{4 \pi} \right)^{2/3}$ and the total pressure of the gas reads $P = P_0 + \tfrac{Qn_0^2}{6}\left( \tfrac{3V}{4 \pi} \right)^{2/3}$, with $P_0 = nK_B T$ the ideal gas contribution. Correcting the pressure in the form of a polytropic equation of state as investigated above, $P = C_\gamma n^{\gamma}$, yields
\begin{equation}\label{eq:correction}
\gamma = 1 + \frac{2/3 \xi}{1+\xi}, \quad \text{with} \quad \xi = \frac{1}{15}\left(\frac{3N}{4 \pi n_0}\right)^{2/3} \frac{\Omega^{5/3}}{a_\gamma^2}.
\end{equation}
A detailed derivation of this result can be found in the Supplemental Material \cite{supp}. Note that, by assuming a constant density distribution, we are overestimating the correction of the polytropic exponent. For that reason, we make the substitution $N \rightarrow N^{\text{eff}} = \alpha N$, with $\alpha$ expected to be close to unit, $\alpha \lesssim 1$. In fact, allowing $\alpha$ to be a free fitting parameter yields $\alpha \simeq 0.8$ as expected. We finally obtain a universal form for the correction of the polytropic exponent, which agrees very well with our theory - see Fig. (\ref{fig_correction}).  
\par
In conclusion, we experimentally determined the equation of state of a gas of cold atoms in large atomic traps by fitting the density profiles with the solution of a generalized Lane-Emden equation describing the hydrostatic equilibrium of the gas. By explicitly evaluating the energy associated with the effective electrostatic interaction, we were able to explain how the polytropic exponent depends of the mean-field potential of the atoms, which in its turn results from the exchange of scattered photons by the atoms. Our findings constitute a strong quantitative evidence of the fact that a laser-cooled gas can effectively simulate and behave like a weakly correlated one-component plasma. Recently, a Debye-H\"uckel approach was introduced in the context of spin ice, where effective magnetic monopoles interact under a mutual Coulombic force \cite{ice}, thus constituting another example where an effective plasmonic behaviour can be induced. The Lane-Emden formalism was, to the best of our knowledge, for the first time applied outside the context of astrophysics. The results obtained here pave the way to the subsequent experimental investigation of more exotic plasma-like processes in non degenerate cold gases. We have previously introduced the possibility of observing effects like phonon-lasing \cite{phonon_laser}, classical rotons \cite{rotons}, plasmon modes and Tonks-Dattner resonances, \cite{tito_2008, livro}, photon bubbles \cite{bubbles}, the dynamical Casimir-effect \cite{Casimir} and twisted excitations carrying orbital angular momentum \cite{twisted}. Multiple scattering should also play a role in the context of opto-mechanical instabilities in cold matter, recently proposed \cite{opto1} and observed \cite{opto2}. We conclude by referring to the close relation between the system investigated here and astrophysical processes involving trapped plasmas \cite{book_astro, astroplasmas}. Being able to achieve mimicking conditions in cold atoms laboratory experiments, with a great degree of control and tunability on the interactions, offers an ideal test bench to investigate astrophysical problems.
\par
We thank R. Kaiser for helpful initial discussions during the set up of the MOT. JR acknowledges the financial support of FCT - Funda\c{c}\~{a}o da Ci\^{e}ncia e Tecnologia through the grant number SFRH/BD/52323/2013. HT thanks the support from Funda\c{c}\~{a}o para a Ci\^{e}ncia e a Tecnologia (Portugal), namely through programmes PTDC/POPH and projects UID/Multi/00491/2013, UID/EEA/50008/2013, IT/QuSim and CRUP-CPU/CQVibes, partially funded by EU FEDER, and from the EU FP7 projects LANDAUER (GA 318287) and PAPETS (GA 323901). 
\subsection{Supplemental Material}
The purpose of this Supplemental Material is to guide the reader through the derivation of the correction of the polytropic exponent, $\gamma$. This parameter accounts for the deviation from an ideal gas ($\gamma = 1$) induced by multiple scattering of photons \cite{sesko_2, dalibard}. The atoms behave as if they possess an effective electrical charge $q=\sqrt{\epsilon_0 Q}$, with $Q = (\sigma_R - \sigma_L)\sigma_L I_0 / c$ \cite{Q}, $I_0$ the total intensity of the beams and $c$ is the speed of light. Here, $\sigma_R$ and $\sigma_L$ represent the emission and absorption cross sections, respectively \cite{walker_1990}. The induced collective interaction has been previously explored \cite{tito_2008, livro}.
\par
Let us begin by determining the electrostatic potential of the system, which is known to satisfy the Poisson equation
\begin{equation}\label{poisson1}
\nabla^2 \phi\left(r\right) = - \frac{1}{\epsilon_0} q n \left(r\right).
\end{equation}
We approximate the density distribution in the cloud by a \textit{water-bag} profile - remember the multiple scattering regime discussed in the main text - corresponding to a constant density $n_0$ spread over a radial extent of radius $R$, i.e. $n\left(r\right) = n_0 \theta \left(r - R \right)$, with $n_0 = 3m\omega_0^2 / Q$ and $R = \left( \frac{3N}{4 \pi n_0} \right)^{1/3}$. This approximation allow us to keep the analysis tractable and derive an analytical correction for $\gamma$. We shall then compute the solutions for the Poisson equation in two different regions. For outside the cloud, $r > R$, Eq. (\ref{poisson1}) reads, in spherical coordinates
\begin{equation}\label{poisson2}
\left( \frac{\partial^2}{\partial r^2} + \frac{2}{r} \frac{\partial}{\partial r} \right) \phi\left( r \right) = 0
\end{equation}
which admits solutions in the form $\phi\left( r \right) = \frac{A}{r} + B$. Assuming that the potential vanishes at infinity, $\phi\left( r \rightarrow \infty \right) = 0$ results in $B=0$. Gauss theorem allows us to write $A = \frac{q_{\text{T}}}{4 \pi \epsilon_0}$ with $q_{\text{T}}$ the total charge of the system, $q_{\text{T}} = \frac{4}{3} \pi q n_0 R^3$. We then have, for $r> R$, $\phi\left( r \right) = \frac{q n_0 R^3}{3 \epsilon_0 r}$. Let now turn to the region inside the cloud, $r \leq R$, where the corresponding Poisson equation reads
\begin{equation}\label{poisson3}
\left( \frac{\partial^2}{\partial r^2} + \frac{2}{r} \frac{\partial}{\partial r} \right) \phi\left( r \right) = -\frac{q n_0}{\epsilon_0}
\end{equation}
In this case the solution are of the form $ \phi\left( r \right)= A' r^2 + B'$. Substituting in Eq. (\ref{poisson3}) results in $A' = - \frac{qn_0}{6 \epsilon_0}$. The integration constant $B'$ is determined by the continuity of the potential $\phi(r)$ in the boundary of the two regions, i.e $\phi(R^{-}) = \phi(R^{+})$. Finally, we can write the electrostatic potential inside the cloud as
\begin{equation}\label{phi1}
\phi\left(r\right) = \frac{qn_0}{6 \epsilon_0} \left( R^2 - r^2 \right) + \frac{q n_0 R^2}{3 \epsilon_0}
\end{equation}
The next step is the evaluation of the effective electrostatic energy, determined by
\begin{equation}\label{int1}
U_{\text{C}} = \frac{1}{2} \int_V \phi \left( r \right) q n\left(r\right) dV
\end{equation}
Introducing again the \textit{water-bag} density profile yields the results $U_{\text{C}} = \frac{4}{15} \pi Q n_0^2 R^5$ or, equivalently, $U_{\text{C}} = \frac{1}{5}\left( \frac{3}{4\pi} \right)^{2/3} \frac{QN^2}{V^{1/3}}$ in terms of the volume and the number of particles in the system, which will be useful in the next steps. We now wish to evaluate the pressure in the cloud, which encompasses the contributions from the ideal gas part and the effective electrostatic interaction, $P = P_0 + P_{\text{C}}$, with $P_0 = k_B T n = k_B T N / V$ and $P_{\text{C}}$ determined by
\begin{equation}
P_{\text{C}} = - \left(\frac{\partial U_{\text{C}}}{ \partial V}\right)_N \text{,}
\end{equation}
since the electrostatic energy doesn't dependent on the temperature. We then have $P_{\text{C}} = \frac{1}{15} \left( \frac{3}{4 \pi} \right)^{2/3} Q N^2 V^{-4/3}$ or, equivalently,  $P_{\text{C}} = \frac{1}{15}Qn_0^2R^2$. The total pressure in the system is given by
\begin{equation}\label{pressure}
P = \frac{k_BT N}{V} + \frac{1}{15} \left( \frac{3}{4 \pi} \right)^{2/3} \frac{Q N^2}{V^{4/3}}
\end{equation}

We now wish to establish an equivalence between the former equation of state and a polytropic-like one, in the form $P=C_{\gamma} n^{\gamma}$, as in the Lane-Emden derivation. With that in mind, we can write
\begin{equation}
C_\gamma n_0^\gamma = k_B T n_0 + \frac{Q n_0^2 R^2}{15}
\end{equation}
 or, equivalently, dividing by $k_B T n_0$
\begin{equation}\label{eqf}
\frac{C_\gamma}{k_B T} n_0^\epsilon = 1 + \frac{Qn_0R^2}{15 k_B T}
\end{equation}
where we defined $\epsilon = \gamma -1 $. Note that we can rewrite this last expression in terms of the parameter of the model introduced earlier, namely the effective plasma frequency $\Omega = \frac{Qn_0}{3 m \omega_0^2}$ and the scaling factor $a_\gamma^2 = \frac{C_\gamma}{3 m \omega_0^2} n_0^\epsilon$. Simple mathematical manipulation of Eq.(\ref{eqf}) finally yields
\begin{equation}
\gamma  = 1 + \frac{2/3 \xi}{\xi + 1}
\end{equation}
with $\xi$ an adimensional universal parameter defined as $\xi = \frac{1}{15}\left( \frac{3N}{4 \pi n_0} \right)^{2/3} \frac{\Omega}{a_\gamma^2}$, where we use the total number of atoms as $N=\frac{4}{3} \pi n_0 R^3$.

\bibliography{Equation_of_State_ArXiv.bib}

\end{document}